\documentclass[%
 reprint, 
nofootinbib,
nobibnotes,
 amsmath,amssymb,
 aps, 
prd,
longbibliography
]{revtex4-2}
\usepackage{xcolor}
\usepackage{natbib}

\usepackage{graphicx}
\usepackage{dcolumn}
\usepackage{bm}
\usepackage{hyperref}
\usepackage{enumitem}
\usepackage{footmisc}
\usepackage{aas_macros}

\begin{document}


\title{\textbf{Extracting cosmological information from the abundance of galaxy clusters with simulation-based inference} 
}

\author{\'{I}\~{n}igo Zubeldia$^{1,2}$}
\email{Contact author: inigo.zubeldia@ast.cam.ac.uk}
\author{Boris Bolliet$^{3,2}$}
\author{Anthony Challinor$^{1,2,4}$}
\author{William Handley$^{1,2}$}

\affiliation{$^1$Institute of Astronomy, University of Cambridge, Madingley Road, Cambridge CB3 0HA, UK}

\affiliation{$^2$Kavli Institute for Cosmology, University of Cambridge, Madingley Road, Cambridge CB3 0HA, UK}

\affiliation{$^3$Cavendish Astrophysics, University of Cambridge, Madingley Road,\,\,Cambridge CB3 0HA, UK}

\affiliation{$^4$DAMTP, Centre for Mathematical Sciences, Wilberforce Road, Cambridge, CB3 0WA, UK}

\date{\today}

\begin{abstract}
The abundance of galaxy clusters as a function of mass and redshift is a well-established and powerful cosmological probe. Cosmological analyses based on galaxy cluster number counts have traditionally relied on explicitly computed likelihoods, which are often challenging to develop with the required accuracy and expensive to evaluate. In this work, we implement an alternative approach based on simulation-based inference (SBI) methods that relies solely on synthetic galaxy cluster catalogues generated under a given model. These catalogues are much easier to produce than it is to develop and validate a likelihood. We validate this approach in the context of the galaxy cluster survey of the upcoming Simons Observatory for a setup in which we can also evaluate an exact explicit likelihood. We find that our SBI-based approach yields cosmological parameter posterior means that are within $0.2\,\sigma$ of those obtained with the explicit likelihood and with biases smaller than $0.1\,\sigma$. We also introduce and validate a procedure to assess the goodness of fit using only synthetic catalogues similar to those used for training. This demonstrates, for the first time, that a galaxy cluster number count cosmological analysis can be performed fully without resorting to a likelihood at any stage. Finally, we apply our SBI-based approach to the real \textit{Planck} MMF3 cosmology sample, obtaining cosmological parameter constraints that are within $0.1\,\sigma$ of their likelihood-based counterparts. This constitutes the first SBI-based number count cosmological analysis of a real galaxy cluster catalogue. 
\end{abstract}

\maketitle


\section{Introduction}\label{sec:introduction}

The abundance of galaxy clusters as a function of mass and redshift is a powerful cosmological probe, sensitive to cosmological parameters such as the matter density parameter $\Omega_{\mathrm{m}}$, the amplitude of matter clustering $\sigma_8$, the equation of state of dark energy, and the sum of the neutrino masses (\cite{Battye:2003bm,Allen2011,Weinberg13}). This has been demonstrated over the past two decades in a number of cosmological `cluster number count' analyses using clusters detected in X-ray, optical and millimetre observations (e.g., \cite{Rozo2010,  Hasselfield2013, Planck2014XX, Mantz2015, Bocquet2015, deHaan2016,  Ade2016,  Bocquet2018, Zubeldia2019, Bolliet2019, Costanzi2019, Abdullah2020, Abbot2020, To2021,  Garrel2022, Chaubal2022, Lesci2022, Chiu2023, Sunayama2023,  Fumagalli2024, Bocquet2024b, Ghirardini2024,Lee2025,Zubeldia2024b,DES2025}). With $10^4$--$10^5$ objects, cluster catalogues from current and upcoming observatories such as eROSITA \cite{Merloni2012}, \textit{Euclid} \cite{Euclid2011}, the Vera C. Rubin Observatory (Rubin/LSST; \cite{LSST2012}), SPT-3G \cite{Benson2014}, the Simons Observatory (SO; \cite{SO2019}), and CMB-S4 \cite{Abazajian2016} have the potential to improve the constraints derived from their predecessors significantly, taking advantage of their sheer statistical power and of the synergies between different observations (e.g., using galaxy weak-lensing observations to calibrate X-ray and thermal Sunyaev--Zeldovich (tSZ) mass--observable scaling relations empirically, as in, e.g., \cite{Bocquet2018, Bocquet2024b, Ghirardini2024}).

Thus far, all published cosmological cluster number count analyses have relied on explicitly computing the likelihood of the data. Cluster number count likelihoods are complex objects in which the population of clusters across redshift and one or more mass observables is typically modelled with a Bayesian hierarchical population model. Evaluating these likelihoods often requires the efficient computation of numerous multi-dimensional integrals (see, e.g., \cite{deHaan2016,Bocquet2018,Zubeldia2019,Bocquet2024a,Zubeldia2024,Ghirardini2024}). Developing a likelihood for a given cluster catalogue, and ensuring its required accuracy across the relevant region of parameter space with the constraint of an acceptable evaluation time, have traditionally demanded a very significant investment of resources.

In this work we implement an alternative approach for obtaining cosmological constraints from galaxy cluster catalogues that does not require the explicit use of a likelihood. This approach makes use of recent developments in simulation-based inference (SBI) methods (see \cite{Cranmer2020} for a review), in particular in neural posterior estimation, and relies solely on synthetic cluster catalogues generated under a given model. These catalogues are significantly easier to generate than it is to develop and validate an explicit likelihood, making this SBI-based approach particularly attractive. We validate our approach in the context of an SO-like survey, but expect it to be broadly applicable. We then apply it to a real dataset, the \textit{Planck} MMF3 cosmology galaxy cluster sample \cite{Planck2015XXVI,Ade2016}.

This paper is organised as follows. First, in Sec.\,\ref{sec:methods} we describe our SBI-based approach and validate it for an SO-like survey, also discussing the goodness of fit within the SBI framework.
Next, in Sec.\,\ref{sec:planck} we apply it to the \textit{Planck} MMF3 cosmology sample and in Sec.\,\ref{sec:comparison} we compare our results to previous work. 
We then discuss the advantages of our SBI-based approach in Sec.\,\ref{sec:case} before concluding in Sec.\,\ref{sec:conclusion}. Appendix\,\ref{appendix:a} lists the parameter priors and the constraints that we obtain in our analyses, and Appendix~\ref{appendix:b} discusses the run-time performance of our SBI-based approach relative to the traditional likelihood-based one.





\section{Simulation-based inference for galaxy cluster cosmology}\label{sec:methods}

\subsection{Demonstration scenario: The Simons Observatory cluster survey}


In a galaxy cluster number count cosmological analysis, the input data set consists of a catalogue of galaxy clusters. Each cluster must have at least one of the following: (i) a redshift measurement; and (ii) measurements for one or several mass observables, i.e., cluster observables that scale with cluster mass.

In this work, we test our proposed cosmological inference method in the context of the galaxy cluster catalogue due to be delivered by the Simons Observatory, an upcoming state-of-the-art millimetre observatory. One of the major goals of the SO Collaboration’s cosmological programme is to produce a catalogue of galaxy clusters detected via the thermal Sunyaev--Zeldovich (tSZ) effect and to derive number count cosmological constraints from it. For SO's baseline noise levels and assuming a cosmology consistent with \textit{Planck} CMB constraints, about 16\,000 galaxy clusters are forecast to be detected \cite{SO2019,Zubeldia2024}, with this number expected to double for the full Advanced SO (ASO) experiment \cite{SO2025}.

Here, we consider an SO-like cluster survey covering $40$\,\% of the sky with a catalogue in which every cluster has a measurement for two mass observables: the tSZ signal-to-noise, $q_{\mathrm{obs}}$, and the CMB lensing signal-to-noise, $p_{\mathrm{obs}}$. Every cluster also has a redshift measurement, $z$, yielding a total of three data points per cluster. The catalogue is assumed to be constructed by imposing a selection threshold of $q_{\mathrm{obs}} = 5$. The cosmological information is extracted from the cluster abundance in the $q_{\mathrm{obs}}$--$z$ plane, with the $p_{\mathrm{obs}}$ measurements serving to calibrate the scaling relation between cluster mass and the tSZ signal-to-noise , which cannot be accurately predicted from first principles.

The two mass observables are linked to cluster mass and redshift with a two-layer hierarchical model. In the first layer, the \emph{mean} tSZ signal-to-noise  $\bar{q} (M_{500},z)$ is given by

\begin{equation}\label{eq:SNR_mean}
    \bar{q} (M_{500},z) = \frac{y_0 ( \beta_{\mathrm{SZ}} M_{500},z)}{\sigma_{y_0} (\theta_{500}  (\beta_{\mathrm{SZ}} M_{500},z))},
\end{equation}
where $y_0$ is the cluster's central Compton-$y$ value, $\sigma_{y_0}$ is the cluster detection noise evaluated at the cluster's angular scale $\theta_{500}$, $M_{500}$ is the cluster mass, and the tSZ mass bias is $\beta_{\mathrm{SZ}} = 0.8$.
We relate $y_0$ to mass by

\begin{equation}\label{eq:scalrelsz}
    y_0 = 10^{A_{\mathrm{SZ}}} \left( \frac{ M_{500}}{ 3 \times 10^{14} h_{70}^{-1} M_{\odot}} \right)^{\alpha_{\mathrm{SZ}}} E^2(z) h_{70}^{-1/2},
\end{equation}
where $A_{\mathrm{SZ}}$ parametrises the tSZ signal amplitude and $\alpha_{\mathrm{SZ}}$ its mass dependence. Furthermore, $E(z)$ is the Hubble parameter scaled by the current value, $H_0 = 100h \,\text{km}\,\text{s}^{-1}\,\text{Mpc}^{-1}$, and $h_{70} = h/0.7$. This scaling relation is consistent with the universal pressure profile of \cite{Arnaud2010} and its form follows the tSZ scaling relation of  \cite{Hilton2018}. The detection noise $\sigma_{y_0}$ is computed by applying the tSZ cluster finder \texttt{SZiFi}\footnote{\href{https://github.com/inigozubeldia/szifi}{\texttt{github.com/inigozubeldia/szifi}}} \cite{Zubeldia2022,Zubeldia2023} to SO-like maps from the Websky simulation \citep{Stein2019,Stein2020}; see \cite{Zubeldia2024}, where similar catalogues are used, for further details.

Still in the first layer of the hierarchical model, the \emph{mean} CMB lensing signal-to-noise, $\bar{p} (M_{500},z)$, is given by

\begin{equation}
    \bar{p} (M_{500},z) = \frac{\kappa_0 (\beta_{\mathrm{CMBlens}} M_{500},z)}{\sigma_{\kappa_0} (\theta_{500} (\beta_{\mathrm{CMBlens}} M_{500},z))},
\end{equation}
where $\kappa_0 (M_{500},z)$ is the central value of the cluster's CMB lensing convergence, 
$\sigma_{\kappa_0}$ is the CMB lensing matched-filter noise, and the CMB lensing mass bias is $\beta_{\mathrm{CMBlens}}=0.92$. This expression assumes that the cluster CMB lensing signal has been extracted with a matched-filter approach, as first proposed in \cite{Melin2015} and applied, e.g., in \cite{Ade2016,Zubeldia2019,Zubeldia2020,Huchet2024}. Following \cite{Zubeldia2019,Zubeldia2020}, in order to compute both $\kappa_0$ and the matched-filter noise, we assume that the cluster convergence profile is that of a truncated Navarro--Frenk--White profile (NFW; \cite{Navarro1997}) with a concentration $c_{\mathrm{500}}=3$ and a truncation radius of $5 R_{500}$, with $\beta_{\text{CMBlens}}$ accounting for the bias in the inferred masses due to the mismatch between the assumed cluster profile and the true mean one (see \cite{Zubeldia2020}). We compute the matched-filter noise using the publicly-available SO minimum-variance (temperature+polarisation) quadratic estimator reconstruction noise curve forecast\footnote{\href{https://github.com/simonsobs/so_noise_models/tree/master/LAT_lensing_noise/lensing_v3_1_1/nlkk_v3_1_0_deproj0_SENS1_fsky0p4_qe_lT30-3000_lP30-5000.dat}{\texttt{github.com/simonsobs/so\_noise\_models/blob/master/ \\ LAT\_lensing\_noise/lensing\_v3\_1\_1/nlkk\_v3\_1\_0\_deproj0\_SENS1 \\ \_fsky0p4\_qe\_lT30-3000\_lP30-5000.dat }}}, taking it to be the same for all the clusters in the sample.

The logarithms of the mean tSZ and CMB lensing signals-to-noise are then linked to the logarithms of the \emph{true} tSZ and CMB lensing signals-to-noise through Gaussian intrinsic scatter, with a covariance matrix given by $\sigma_{\mathrm{SZ}}$, $\sigma_{\mathrm{CMBlens}}$, and a correlation coefficient $r = 0$ (i.e., no intrinsic correlation). This log-normal intrinsic scatter accounts for true scatter in the cluster observables (e.g., due to cluster profile variations, triaxiality, and large-scale structure along the line of sight). In the second layer of the model, the set of scaling relations simply exponentiates $\ln q$ and $\ln p$, which are then linked to the observed values ($q_{\mathrm{obs}}$ and $p_{\mathrm{obs}}$, respectively) through uncorrelated Gaussian scatter with unit variance for both observables. 

We note that this two-layer Bayesian population model describing the cluster observables is the same as the one that was adopted in a similar SO context in \cite{Zubeldia2024}.

\subsection{Parameters to be constrained}

We assume a flat $\Lambda$ cold dark matter ($\Lambda$CDM) cosmological model and consider a 10-dimensional parameter space. Five of these parameters are cosmological: $\sigma_8$, which parametrises the amplitude of large-scale matter clustering; the physical cold dark matter matter density parameter, $\omega_{\mathrm{c}}$; the physical baryon matter density parameter, $\omega_{\mathrm{b}}$; the Hubble constant in units of 100\,km\,s$^{-1}$\,Mpc$^{-1}$, $h$; and the spectral index of the primordial scalar perturbations, $n_\mathrm{s}$. The remaining five parameters are: the tSZ cluster amplitude parameter $A_{\mathrm{SZ}}$, the tSZ cluster slope parameter $\alpha_{\mathrm{SZ}}$, the scatter in the tSZ signal-to-noise $\sigma_{\mathrm{SZ}}$, the CMB lensing bias parameter $\beta_{\mathrm{CMBlens}}$, and the scatter in the CMB lensing signal-to-noise, $\sigma_{\mathrm{CMBlens}}$. From a cosmological point of view, they can be thought of as nuisance parameters to be marginalised over.

\subsection{Synthetic data generation}

We generate our synthetic catalogues using the catalogue generator of the publicly available \texttt{cosmocnc} code \cite{Zubeldia2024}. At a given point in parameter space, the catalogue generator first samples cluster mass--redshift pairs from the halo mass function, producing a realisation of the total population of clusters in the Universe down to a lower mass limit $M_{\mathrm{min}} = 5 \times 10^{13} M_{\odot}$. The halo mass function is efficiently evaluated using the \texttt{cosmopower} neural network emulators \cite{Spurio2022,Bolliet2024} as implemented in \texttt{class\_sz} \cite{Bolliet2023b}. The clusters are assigned random locations in the sky, neglecting the effect of cluster clustering; see Sec.\,\ref{sec:case} for how this effect could be included. Then, for each cluster, the scaling relations of the mass observables (in this case, $q_{\mathrm{obs}}$ and $q_{\mathrm{obs}}$) are applied and random scatter is added. This procedure is repeated twice to account for the two layers of scatter in our model, intrinsic and from measurement errors. The catalogue is then obtained by imposing the selection criterion, i.e., by selecting all clusters with  $q_{\mathrm{obs}} > 5$. We refer the reader to Sec.~4.9.1 of \cite{Zubeldia2024} for a more detailed description of the catalogue generator.

With $N_{\mathrm{cluster}} \simeq $\,16\,000 clusters at a \textit{Planck} cosmology \cite{Zubeldia2024} and three data points per cluster, our catalogues are too high-dimensional for current state-of-the-art SBI methods \cite{Cranmer2020}. To address this problem, we compress each catalogue by binning the $q_{\mathrm{obs}}$--$z$ pairs into a $4 \times 5$ grid, defined by five logarithmically spaced bin edges between $q_{\mathrm{obs}} = 5$ and $q_{\mathrm{obs}} = 40$ and six linearly spaced bin edges between $z = 0.01$ and $z = 1.5$. This two-dimensional grid is chosen to encompass most SO clusters for a \textit{Planck} cosmology (see, e.g., Fig.~2 of \cite{Zubeldia2024}). In addition, the CMB lensing signal-to-noise measurements are averaged across the entire catalogue, producing a single stacked measurement $\hat{p}_{\mathrm{stacked}}$. This choice is justified given the very small per-cluster signal-to-noise of this observable (see \cite{Zubeldia2024}). In brief, each initial $3 \times N_{\mathrm{cluster}}$-dimensional data vector is therefore reduced to a 21-dimensional data vector (20 counts in cells and a single stacked mass measurement). Given the comparisons between the binned, unbinned, and unbinned+stacked number count likelihoods presented in \cite{Zubeldia2024}, this data compression procedure is effectively lossless for SO in a $\Lambda$CDM cosmology.

To train our inference method, we generate and compress catalogues for $N_{\mathrm{sim}}=10\,000$ points in parameter space. These points are drawn from the parameter priors we adopt, which are specified in Table~\ref{table} in Appendix~\ref{appendix:a}. In particular, we note that we impose wide uniform priors on the two cosmological parameters that cluster number counts can constrain precisely, $\sigma_8$ and $\omega_{\mathrm{c}}$. Similarly, we impose wide uniform priors on  $A_{\mathrm{SZ}}$ and $\alpha_{\mathrm{SZ}}$, leaving these parameters to be constrained by the CMB lensing data. We impose Gaussian priors on the remaining parameters.

Finally, we generate and compress a set of 100 test catalogues at what we regard as our true point in parameter space. The true parameter values are listed in Table~\ref{table}.

\subsection{Inference methods}


We learn the posterior of our compressed catalogue data using the neural posterior estimation (NPE) method of \cite{Greenberg2019}, as implemented in the \href{https://sbi-dev.github.io/sbi/latest/}{sbi} package \cite{Tejero-Cantero2020}, where it corresponds to the \texttt{NPE\_C} method, leaving the exploration of other SBI methods (e.g., \cite{Papamakarios2019,Alvey2023}) to further work. We perform a single training iteration (i.e., obtain an amortised posterior estimate), adopt the \texttt{maf} density estimator, and terminate the training after 50 epochs of no improvement in the validation loss. We then obtain parameter constraints by drawing $10^5$ samples from the learned posterior. We assess the impact of the learning rate $\eta$ by training with $\eta = 5 \times 10^{-3}$, $5 \times 10^{-4}$, $5 \times 10^{-5}$, and $5 \times 10^{-6}$. We also investigate the impact of the number of training catalogues $N_{\mathrm{sim}}$ by training with $N_{\mathrm{sim}} = 500$, 1\,000, 2\,000, 5\,000, and 10\,000.

\begin{figure*}
\includegraphics[width=0.85\textwidth]{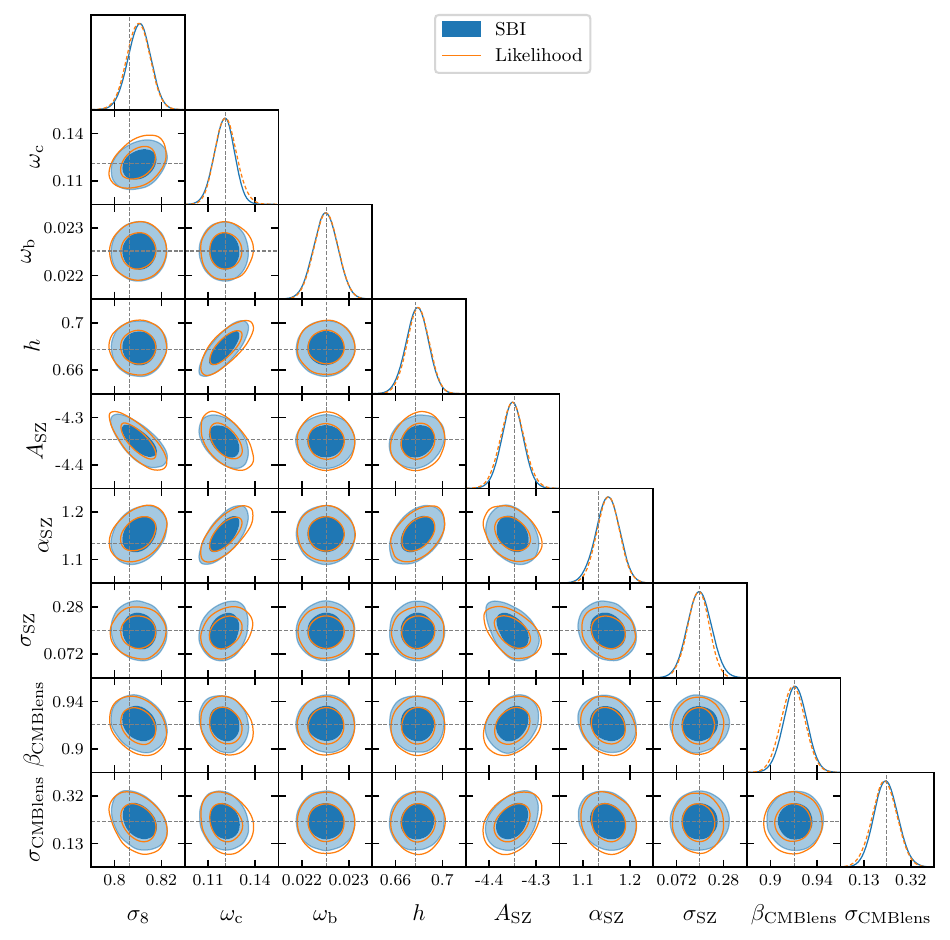}
\caption{\label{fig:constraints} Parameter constraints obtained with our SBI-based approach (blue filled contours) and our likelihood-based approach (orange contours) for our reference SO-like catalogue. The constraints from both methods agree to within $0.2\,\sigma$ for all parameters (see Table\,\ref{table}). The dashed lines indicate the true parameter values.}
\end{figure*}

In addition, we infer the posterior of one of our test cluster catalogues (our `reference test catalogue') following a traditional likelihood-based approach, adopting the same priors as in our SBI analysis. Specifically, we use the \texttt{cosmocnc}  unbinned+stacked likelihood, which provides an almost exact likelihood of our synthetic catalogues. The only approximation is the assumption of Gaussian scatter in the stacked observable, which is highly accurate given  the large number of objects being averaged over \cite{Zubeldia2024}. This likelihood has been validated in a similar SO context (see \cite{Zubeldia2024}). We sample from the posterior using the Markov chain Monte Carlo (MCMC) \texttt{Cobaya} package \cite{Torrado2019,Torrado2021}.

\begin{figure}
\includegraphics[width=0.53\textwidth,trim={5mm 0mm 0mm 10mm},clip]{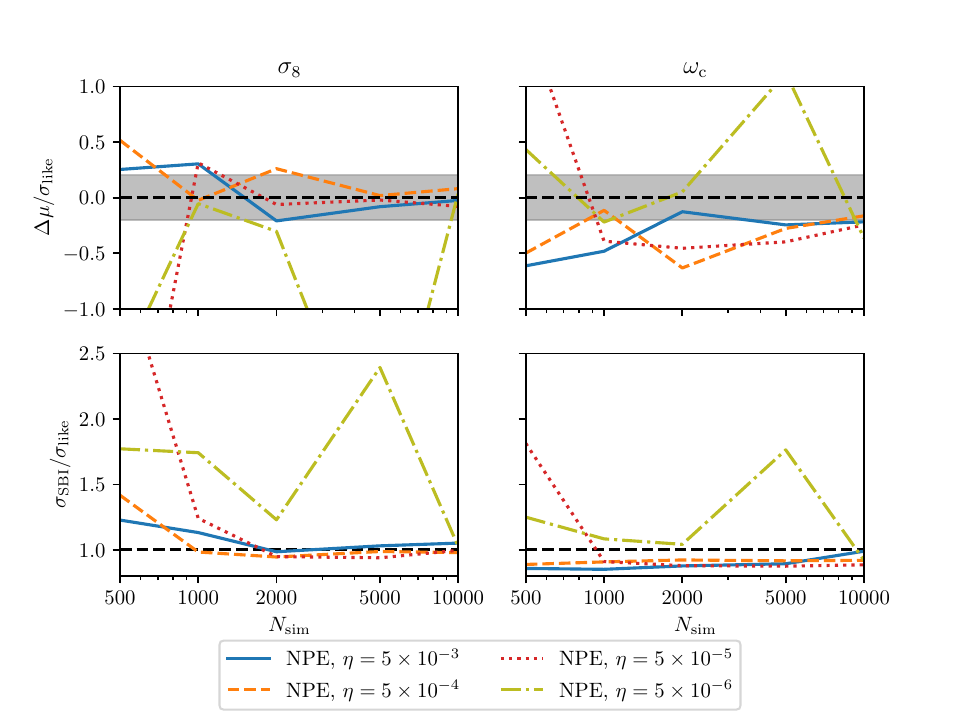}
\caption{\label{fig:convergence} \textit{Upper panels}: Difference in the mean inferred values between our SBI-based and likelihood-based approaches, $\Delta \mu \equiv \mu_{\mathrm{SBI}} - \mu_{\mathrm{like}}$, expressed in units of the standard deviation of the likelihood-based constraint $\sigma_{\mathrm{like}}$, as a function of $N_{\mathrm{sim}}$ and the learning rate $\eta$ for our two cosmological parameters of interest. The shaded region represents a $\pm0.2$\,$\sigma$ deviation from the likelihood constraint. \textit{Lower panels}: Ratio of the standard deviation of the SBI constraint, $\sigma_{\mathrm{SBI}}$, to that of the likelihood constraint, $\sigma_{\mathrm{like}}$.}
\end{figure}

\subsection{Results}\label{sec:results}

Figure~\ref{fig:constraints} shows the parameter constraints obtained for our reference test catalogue following: (i) our SBI approach with our `baseline configuration' ($N_{\mathrm{sim}} = 10\,000$ and $\eta = 5 \times 10^{-4}$; filled blue contours); and (ii) our likelihood-based approach (orange contours). The agreement between the two methods is excellent, with the differences in the mean inferred values $\Delta \mu \equiv \mu_{\mathrm{SBI}} - \mu_{\mathrm{like}}$ being less than 0.2\,$\sigma$ for all parameters (see Table~\ref{table} in Appendix~\ref{appendix:a}). Notably, the difference is of only 0.081\,$\sigma$ for $\sigma_8$. We regard these differences as negligible. We note that ther constraints on $\sigma_8$, $\omega_{\mathrm{c}}$, $A_{\mathrm{SZ}}$, and $\alpha_{\mathrm{SZ}}$ are data driven, whereas those for the remaining parameters are prior driven and therefore they peak very close to the prior peak.

In addition, the two upper panels of Fig.~\ref{fig:convergence} show $\Delta \mu$ in units of the standard deviation of the likelihood-based constraint $\sigma_{\mathrm{like}}$ as a function of $N_{\mathrm{sim}}$ and $\eta$ for our two cosmological parameters of interest, $\sigma_8$ and $\omega_{\mathrm{c}}$ (left and right panel, respectively). The shaded region corresponds to a $\pm0.2$\,$\sigma$ deviation from the likelihood constraint. Defining convergence in our SBI constraints as a change of less than $0.1\,\sigma$  in the constraints on both parameters when doubling $N_{\mathrm{sim}}$, this is achieved for $\eta = 5 \times 10^{-3}$ and $5 \times 10^{-4}$ at $N_{\mathrm{sim}} = 10\,000$. For these configurations, the constraints in both parameters are within 0.2$\,\sigma$ of their likelihood-based counterparts. Furthermore, for $N_{\mathrm{sim}} > 5\,000$ and $\eta = 5 \times 10^{-3}$, $5 \times 10^{-4}$, and $5 \times 10^{-5}$, convergence of the derived constraints within 0.2\,$\sigma$ of each other is observed. A learning rate of $5 \times 10^{-6}$, however, generally results in poorer performance. We note that due to the stochasticity in the training process (random initialisation of the neural network and batch gradient descent), $\Delta \mu/\sigma_{\text{like}}$ and $\sigma_{\text{SBI}}/\sigma_{\mathrm{like}}$ (see below) do not change monotonically with $N_{\text{sim}}$.

The lower panels of Fig.~\ref{fig:convergence} show the ratio between the standard deviation of the SBI constraint, $\sigma_{\mathrm{SBI}}$, and that of the likelihood-based constraint, $\sigma_{\mathrm{like}}$, for the same two cosmological parameters. For $\sigma_{8}$, the standard deviations agree within 5\,\% for $\eta = 5 \times 10^{-3}$, $5 \times 10^{-4}$, and $5 \times 10^{-5}$ when $N_{\mathrm{sim}} > 2\,000$, while for $\omega_{\mathrm{c}}$ the agreement is within 10\,\%.

Finally, we perform a self-standing validation of our SBI approach by obtaining SBI parameter constraints for our 100 test catalogues evaluating the posterior obtained with our baseline configuration, which we recall is amortised. Averaging the parameter posterior mean values across all test catalogues, we find biases of less than $0.1\,\sigma$ for all 10 parameters (see Table~\ref{table} in Appendix~\ref{appendix:a}). In particular, for $\sigma_{8}$ and $\omega_{\mathrm{c}}$ no biases are detected at levels of $0.086\,\sigma$ and $0.063\,\sigma$, corresponding to non-detections at the 0.051\% and 0.27\% levels, respectively.

\subsection{Goodness of fit}\label{sec:goodness}

Synthetic data generators naturally enable the evaluation of the goodness of fit of an analysis. This can be achieved by generating a set of data vectors at the `best-fit' point in parameter space. The full data vector distribution at that point can then be mapped and the consistency of the observed data vector with it can be quantified. Alternatively, goodness-of-fit summary statistics can be computed.

Here, we assess the goodness of fit of the SBI-based analysis of our reference SO-like catalogue by generating and compressing 100 synthetic catalogues at the `best-fit' point in parameter space (our `best-fit' catalogues) and evaluating two statistics. These statistics are: (i) the modified Cash statistic of \cite{Kaastra2017}, $\hat{C}$, which can be used to assess the goodness of the fit in the number counts by comparing its value to its predicted mean and standard deviation (see, e.g., \cite{Bocquet2018,Zubeldia2024}); and (ii) the value of the stacked observable, $\hat{p}_{\mathrm{stacked}}$. We take the best-fit point to be specified by the posterior means, noting that in other scenarios in which the posterior is more non-Gaussian, other more suitable points, e.g., the maximum a posteriori point, can easily be chosen, as the posterior can be evaluated at any point in parameter space. Alternatively, in a fully Bayesian approach, the best-fit statistics could be evaluated for synthetic catalogues generated at samples from the full posterior distribution \cite{Wietersheim2025}.

In order to compute $\hat{C}$ and its expected value and standard deviation, a theoretical prediction for the number count in each cell is required. We obtain this prediction by averaging the counts over our 100 best-fit catalogues. Doing this, we obtain an observed value of $\hat{C} = 23.05$ and a predicted mean and standard deviation of  $\bar{C}_{\mathrm{sim}} = 20.19$ and $\sigma_{C,\mathrm{sim}} = 6.35$, respectively. We can therefore conclude that the best-fit prediction is a good fit to the data, as $\hat{C}$ is consistent with its predicted mean $\bar{C}_{\mathrm{sim}}$. For comparison, we also evaluate this statistic using a simulation-free prediction for the number counts produced with \texttt{cosmocnc}, finding $\hat{C} = 21.64$, $\bar{C}_{\mathrm{theory}} = 20.18$, and $\sigma_{C,\mathrm{theory}} = 6.36$, in good agreement with the simulation-derived values. We note that, in addition to evaluating this statistic, the difference between the (simulation- or theory-based) predicted and observed number counts could also be inspected visually.

On the other hand, the observed value of the stacked observable is $\hat{p}_{\mathrm{stacked}} = 0.3082$. We estimate the mean and standard deviation of $\hat{p}_{\mathrm{stacked}}$ by computing its sample mean and standard deviation across the 100 best-fit catalogues. We find $\bar{p}_{\mathrm{stacked,sim}} = 0.3105 \pm 0.0009$ (empirical standard deviation) and $\sigma_{{p}_{\mathrm{stacked}},\mathrm{sim}} = 0.0085$, in agreement with with the observed value. For comparison, we also compute these two quantities with \texttt{cosmocnc} (see \cite{Zubeldia2024}), finding $\bar{p}_{\mathrm{stacked,theory}} = 0.3106$ and $\sigma_{{p}_{\mathrm{stacked}},\mathrm{theory}} = 0.0080$, in excellent agreement with the simulation-derived values.

These results demonstrate that the goodness of fit of an SBI-based cluster number count analysis that includes both counts and stacked observables can be quantified using the same synthetic catalogue generator used to produce the training catalogues. This implies that our SBI-based approach can be successfully applied to data in a fully self-contained way, without the need to resort to an explicit likelihood at any point in the analysis.

\section{Application to real data from \textit{Planck}}\label{sec:planck}

\subsection{The \textit{Planck} cluster catalogue and model}

We apply our SBI-based approach to the cosmology sample of the \textit{Planck} MMF3 galaxy cluster catalogue, which is publicly available on the Planck Legacy Archive\footnote{\href{https://pla.esac.esa.int}{pla.esac.esa.int}} (PLA). This sample consists of 439 clusters detected through their tSZ signature using data from the \textit{Planck} experiment \cite{Planck2015XXVI}. It was used to derive cosmological constraints in the 2015 cluster number count analysis carried out by the \textit{Planck} Collaboration (\cite{Ade2016}; hereafter, the `official \textit{Planck} analysis'), and has subsequently been reanalysed in several studies (e.g., \cite{Zubeldia2019,Salvati2022}).

In the MMF3 cosmology sample, every cluster has a tSZ signal-to-noise measurement, $q_{\text{obs,\textit{Planck}}}$, and a redshift measurement, $z$ (the latter except for 6 clusters). The sample is constructed by imposing a signal-to-noise threshold of 6. This setup is thus very similar to the SO-like one considered in Sec.\,\ref{sec:methods}. 

Here, we adopt the same mass--observable model that was assumed in the official \textit{Planck} analysis, which is also very similar to that of the SO tSZ signal-to-noise adopted in Sec.\ref{sec:methods}. In this model, $q_{\text{obs,\textit{Planck}}}$ is linked to the cluster mass and redshift with a two-layer hierarchical model. In the first layer, the mean signal-to-noise, $\bar{q}_{\text{\textit{Planck}}} (M_{500},z)$, is given by 
\begin{equation}\label{eq:SNR_mean_planck}
    \bar{q}_{\text{\textit{Planck}}} (M_{500},z) = \frac{\bar{Y}_{500} ( (1-b) M_{500},z)}{\sigma_{Y} (\theta_{500}  ((1-b) M_{500},z))},
\end{equation}
where $\bar{Y}_{500} ( (1-b) M_{500},z)$ is the cluster's mean integrated Compton-$y$ value, $\sigma_Y$ is the detection noise (which depends on the sky position and is available on the PLA), and $1-b$ is the tSZ mass bias (also known as the `hydrostatic mass bias'). This scaling relation is equivalent to that of the SO signal-to-noise, shown in Eq.\,\ref{eq:SNR_mean}, the difference being that here, following \cite{Ade2016}, it is parametrised in terms of the integrated Compton-$y$ signal $\bar{Y}_{500}$, instead of the cluster's central Compton-$y$ signal $y_0$. In order to keep the same notation as in the official \textit{Planck} analysis, we denote the tSZ mass bias with $1-b$ rather than $\beta_{\mathrm{SZ}}$ as in Sec.\,\ref{sec:methods}. The mean integrated Compton-$y$ value, $\bar{Y}_{500} (M_{500},z)$, is in turn given by
\begin{multline}
\bar{Y}_{500} ( M_{500},z) = Y_{\star} (h/0.7)^{-2+\alpha} \\ \times \left( \frac{M_{500}}{6 \times 10^{14} M_{\odot}} \right)^{\alpha}   \left( \frac{D_{\mathrm{A}} (z)}{10^{-2} \,\mathrm{Mpc}} \right)^{-2} E^{\beta} (z),    
\end{multline}
where $Y_{\star}$ parametrises the tSZ signal amplitude (analogous to $A_{\mathrm{SZ}}$ in Sec.\,\ref{sec:methods}), $\alpha$ its mass dependence (analogous to $\alpha_{\mathrm{SZ}}$), and $\beta$ its redshift dependence, which, following \cite{Ade2016}, we fix to $\beta=0.66$.

The logarithm of the mean tSZ signal-to-noise is then linked to the logarithm of the true tSZ signal-to-noise $\ln q$ through Gaussian intrinsic scatter with a standard deviation of $\sigma_{\ln Y}$ (analogous to $\sigma_{\mathrm{SZ}}$ in Sec.\,\ref{sec:methods}). In the second layer of the model, $\ln q$ is exponentiated and linked to the observed value $q_{\mathrm{obs},Planck}$ through Gaussian scatter with unit variance, which accounts for observational noise.

As in Sec.\,\ref{sec:methods}, we compress the cluster sample by binning the $q_{\text{obs,\textit{Planck}}}$--$z$ pairs into a $4 \times 5$ grid defined by five logarithmically spaced bin edges between $q_{\mathrm{obs}}=6$ and $q_{\mathrm{obs}}=40$ and six linearly spaced bin edges between $z=0.01$ and $z=1$, spanning all but one of the clusters in the MMF3 cosmology sample. This procedure compresses the cluster sample into a 20-dimensional data vector. Our grid is coarser than that used in the official \textit{Planck} analysis, but we do not expect this to cause any significant loss of constraining power given the smoothness of the $q_{\text{obs,\textit{Planck}}}$--$z$ distribution.

\begin{figure*}
\includegraphics[width=0.85\textwidth]{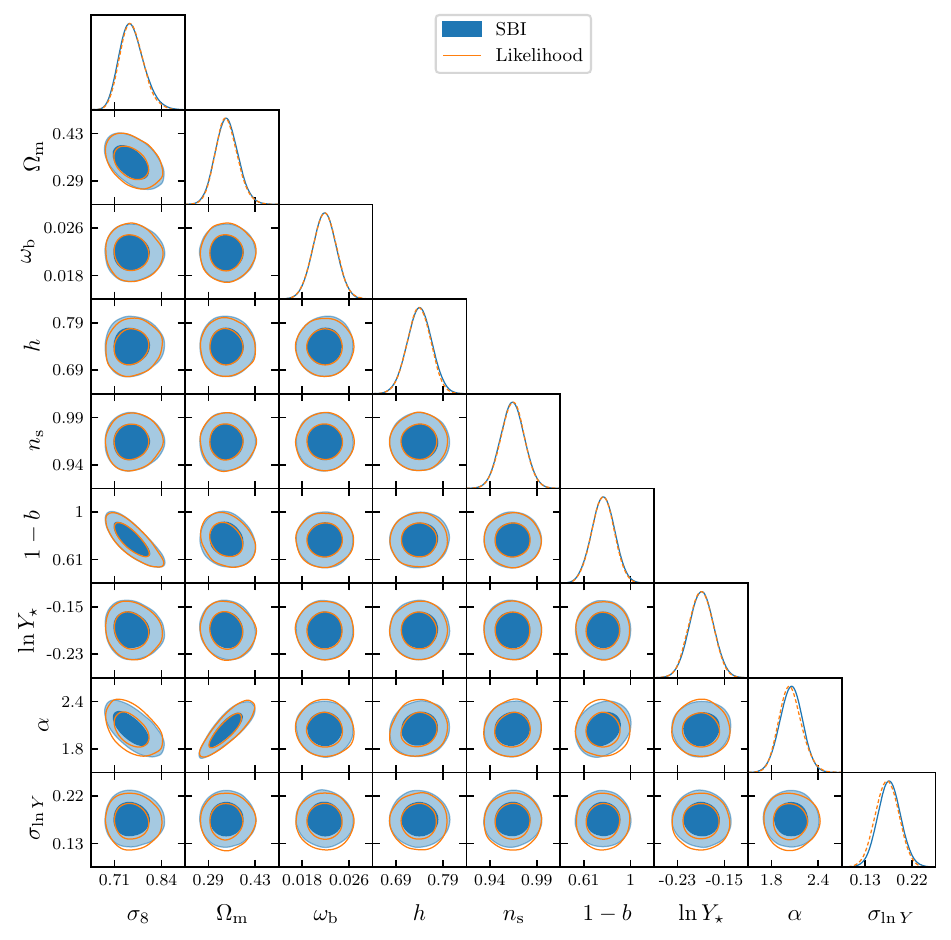}
\caption{\label{fig:constraints_planck} Parameter constraints for the \textit{Planck} MMF3 cosmology sample (real data), obtained with both our SBI-based approach (filled blue contours) and our likelihood-based approach (orange contours). The agreement between both methods is excellent, with differences smaller than 0.07\,$\sigma$ for the posterior means of all parameters (see Table\,\ref{table2}) and with the mildly non-Gaussian shapes of the two-dimensional posteriors being remarkably similar. We emphasise that the adopted priors (including those on $h$ and $1-b$) were chosen in order to have the same setup as in one of the four main analyses in the \textit{Planck} Collaboration 2015 cluster number count paper \cite{Ade2016}.
}
\end{figure*}

\subsection{Cosmological constraints}

As a demonstration of our SBI-based approach on real data, we set out to reproduce the cosmological constraints of the official 2015 \textit{Planck} analysis of \cite{Ade2016}. For simplicity, among their four main analyses, we consider their CCCP+$H_0$+BBN analysis, as the other three involve combining the cluster catalogue with baryon acoustic oscillation (BAO) data. In this analysis, the parameter space is nine-dimensional and is spanned by: $\sigma_8$, $\Omega_{\mathrm{m}}$ (the matter density parameter), $\omega_{\mathrm{b}}$, $h$, $n_{\mathrm{s}}$, $1-b$, $\log Y_{\mathrm{\star}}$, $\alpha$, and $\sigma_{\ln Y}$. Here, we explore the same parameter space and adopt the same priors as in \cite{Ade2016}, which can be found in Table\,\ref{table2} of Appendix\,\ref{appendix:a}. Notably, the weak-lensing-derived prior on the tSZ mass bias is $1-b = 0.780 \pm 0.092$ (mean and standard deviation; from the Canadian Cluster Comparison Project, CCCP, \cite{Hoekstra2015}), and the supernova-derived prior on the Hubble constant is $H_0 = 73.8 \pm 2.4$\,km\,s$^{-1}$\,Mpc$^{-1}$ \cite{Riess2011}. We emphasise that these priors are adopted solely for the purpose of reproducing the results of \cite{Ade2016}.

We apply the SBI-based approach described and validated in Sec.~\ref{sec:methods} to the compressed MMF3 sample, generating and compressing 5\,000 synthetic \textit{Planck} catalogues. We note that these catalogues are generated properly accounting for the sky dependence of the detection noise. We adopt a learning rate of $\eta=5\times 10^{-4}$ and stop the training after 50 epochs of no improvement in the validation loss.

We also infer the posterior of the compressed \textit{Planck} data vector using \texttt{cosmocnc}'s binned likelihood, sampling from the posterior with \texttt{Cobaya}. This likelihood is a Poisson likelihood that assumes the same model as the one adopted to generate the catalogues used in the SBI-based approach. Therefore, it is expected to deliver the same posterior as the SBI-based approach. It is also equivalent to the `two-dimensional' likelihood of the official \textit{Planck} analysis (apart from the different binning resolution).

Figure~\ref{fig:constraints_planck} shows the parameter constraints obtained from the MMF3 cosmology sample using our SBI-based approach (solid blue contours) and the likelihood-based approach (orange contours). The agreement between the two methods is excellent, noting that the mildly non-Gaussian shapes of the two-dimensional posteriors are remarkably similar (see also Table\,\ref{table2}, which lists the parameter constraints for both methods and their differences). For our SBI-based approach, we constrain $\sigma_8 =  0.756 \pm 0.034$, $\Omega_{\mathrm{m}} = 0.342 \pm 0.035$, and $\alpha = 2.05 \pm 0.13$ (mean and standard deviation), noting that the remaining parameters are prior driven (see Table\,\ref{table2}). These mean values are $0.01$, $0.023$, and $0.063\,\sigma$ away from their likelihood-based counterparts, which constitutes a remarkable agreement. We also find $\sigma_8 (\Omega_{\mathrm{m}}/0.3)^{0.3} = 0.777 \pm 0.034$, with both approaches yielding the same value.

Our constraints on $\sigma_8$ and $\Omega_{\mathrm{m}}$ are consistent within $1\sigma$ with those reported in the 2015 \textit{Planck} paper for the same CCCP+$H_0$+BBN analysis setup \cite{Ade2016}, for which $\sigma_8 =  0.78 \pm 0.04$ and $\Omega_{\mathrm{m}} = 0.31 \pm 0.04$. There is better agreement for $\sigma_8 (\Omega_{\mathrm{m}}/0.3)^{0.3}$, for which \cite{Ade2016} report $\sigma_8 (\Omega_{\mathrm{m}}/0.3)^{0.3} = 0.772 \pm 0.034$, 0.15$\sigma$ away from our constraint. Given that the constraints in \cite{Ade2016} are obtained for the same data, model, and priors as in our analysis, the observed differences are likely due to differences in implementation. Since the goal of this exercise is to demonstrate that our SBI-based approach can be applied to real data (and yields constraints that are virtually identical to those obtained in our own likelihood-based analysis), we do not investigate these differences further here.

\section{Comparison to previous work}\label{sec:comparison}

The use of SBI in galaxy cluster cosmology was first proposed, as a proof of concept, in \cite{Ishida2015}, in which the Approximate Bayesian Computation (ABC) method was applied. More recently, \cite{Tam2022} applied both ABC and the \texttt{pydelfi} algorithm \cite{Alsing2019} to optically-selected synthetic cluster catalogues, assessing their performance to the $1$\,$\sigma$ level with one test catalogue. In addition, \cite{Kosiba2025} applied the Sequential Neural Posterior Estimator (SNPE) of \cite{Papamakarios2016} to X-ray-selected synthetic catalogues, finding no evidence for biases in cosmological parameters as the constraining power of the test catalogue was increased. Finally, \cite{Reza2024} applied the same SNPE method to optically-selected synthetic catalogues, finding no biases in the relevant cosmological parameters to the $1$\,$\sigma$ level and consistency with a likelihood-based approach similarly to the $0.5$--$1$\,$\sigma$ level. In addition, they also applied, for the first time, an SBI approach to synthetic cluster catalogues produced using halo catalogues from a cosmological simulation (specifically, from the Quijote simulations; \cite{Villaescusa-Navarro2020}).

The current paper takes the work presented in \cite{Ishida2015,Tam2022,Kosiba2025,Reza2024} further in several ways. First, we constrain the biases of all the parameters to less 0.1\,$\sigma$ in the highly demanding context of SO, improving upon what has been generally demonstrated previously (with the exception of \cite{Kosiba2025}, where no biases were found when the constraining power of the test catalogue was increased by a very large amount, noting that the comparison is difficult given that \cite{Kosiba2025} do not quantify the parameter biases as we do). We also find agreement between our SBI and likelihood-based constraints to less than $0.17$\,$\sigma$ for SO and less than $0.07$\,$\sigma$ for real \textit{Planck} data, improving upon what is reported in \cite{Reza2024} (the only other analysis carrying out such a comparison) for their synthetic halo-mass-function-based catalogues (agreement at the $0.5$--$1$\,$\sigma$ level). In contrast, unlike in \cite{Reza2024}, here we do not attempt to use halo catalogues from cosmological simulations, leaving the exploration of this more realistic avenue to further work. We also apply a modern neural-estimation-based method to a tSZ catalogue for the first time. In addition, we propose and validate, for the first time, an approach to quantify the goodness of fit of an SBI-based cluster number count analysis that relies solely on synthetic catalogues; goodness of fit is not discussed in \cite{Ishida2015,Tam2022,Kosiba2025,Reza2024}. Finally, and importantly, the application of our SBI-based approach to the \textit{Planck} MMF3 cosmology sample constitutes, to the authors' knowledge, the first application of SBI to carry out a cluster number count analysis of a real galaxy cluster catalogue.

\section{The case for SBI-based galaxy cluster cosmology}\label{sec:case}

The results presented in Sec.~\ref{sec:methods} demonstrate that our SBI-based approach for galaxy cluster number count cosmological inference is able to deliver unbiased cosmological constraints in the highly demanding context of SO and provides a natural way to quantify the goodness of fit. In addition, in Sec.~\ref{sec:planck} we have demonstrated that it can be successfully applied to real data. All of this is achieved without the need to develop and implement an explicit likelihood, relying solely on synthetic cluster catalogues.


In the two specific setups we considered we already had an explicit likelihood implementation. However, we argue that even when a likelihood can be formulated and implemented, our SBI-based approach offers several significant advantages, as follows.

\begin{enumerate}[leftmargin=12pt]
\item Cluster number count likelihoods are usually very complex, often involving a large number of multi-dimensional integrals to marginalise over the latent variables in the Bayesian hierarchical population models describing the cluster catalogues (see, e.g., \cite{deHaan2016,Bocquet2018,Zubeldia2019,Bocquet2024a,Zubeldia2024,Ghirardini2024}). Developing and validating such likelihoods requires substantial investment, often taking months to years. In contrast, generating synthetic catalogues, as done in this work, is significantly easier, requiring only efficient sampling from the halo mass function and random number generation to account for the scatter in the cluster observables.
    
\item The accuracy of cluster number count likelihoods can degrade in certain regions of parameter space due to convergence issues in the likelihood integrals. In contrast, provided that the halo mass function is sampled accurately (which is much simpler than computing likelihood integrals), synthetic catalogues are \emph{exact} realisations of the assumed model, ensuring uniform accuracy across the parameter space.

\item The complexity of the model describing the cluster observables can significantly impact how efficiently a likelihood can be computed. Challenges include, e.g., correlated scatter between the mass observables and mass-dependent scatter (see, e.g., \cite{Zubeldia2024}). Synthetic catalogue generation remains unaffected by such complexities\footnote{With the exception of non-Gaussian scatter (not including log-normal scatter, as it is Gaussian in the logarithm of the variable), for which a tailored random number generator would have to be used.}: once the scaling relations and the scatter parameters are specified, the same generator can be used for any catalogue. Moreover, the generation time is not affected by correlated scatter (see Appendix\,\ref{appendix:b}).

\item Evaluating cluster number count likelihoods can be computationally expensive. For instance, the \texttt{cosmocnc} likelihood used in this work takes approximately 12\,s per evaluation for our SO-like catalogue with two mass observables, neglecting correlated scatter. More mass observables and larger catalogues, both typical scenarios in upcoming experiments, will only exacerbate this problem. Furthermore, traditional MCMC methods offer limited parallelisation. In contrast, while generating synthetic catalogues can also be computationally demanding (particularly for accurate sampling of the halo mass function), the process scales minimally with the number of observables (see  Appendix~\ref{appendix:b}) and is trivially parallelisable \textit{en masse}, as the catalogues are generated independently from each other. As shown in Appendix~\ref{appendix:b}, for our SO setup, the SBI analysis achieves a comparable total CPU time but significantly lower wall-clock time than the likelihood-based approach.
\end{enumerate}

An SBI-based approach along the lines of the one proposed here becomes even more advantageous when an accurate likelihood is either computationally prohibitive or simply intractable. In cluster cosmology, this situation can arise when sample variance due to cluster clustering cannot be neglected, making the Poissonian likelihood used here inaccurate (see, e.g., \cite{Hu2003}). While accounting for sample variance in a binned likelihood with a single mass observable is relatively straightforward (see, e.g., \cite{Rozo2010}), doing so in more complicated setups, e.g., when additional mass observables are used for mass calibration (as in this work for the SO-like survey), is significantly more challenging~\cite{Payerne2023}. However, incorporating the effect of sample variance into synthetic cluster catalogues is a simpler problem. This could be achieved, e.g., by sampling from a density-field-modulated halo mass function or by directly using cluster catalogues from cosmological simulations  (e.g., \cite{Villaescusa-Navarro2020,Bayer2024}). These cluster catalogues could then be used to generate synthetic catalogues by applying the cluster observable model to them (see, e.g., \cite{Reza2024}). Further realism could be achieved by using catalogues produced by applying the cluster finder used to obtain the real catalogue to simulated sky maps. Doing this would account for sample variance due to cluster clustering in the inference process, as well as for other complications such as blending of several clusters into a single detection. It would also allow to drop simplifying assumptions in the model, such as log-normal intrinsic scatter and power-law scaling relations. We leave the exploration of these promising approaches to future work.

Moreover, SBI naturally facilitates the combination of cluster catalogues with correlated datasets, such as cosmic shear or tSZ power spectrum data, provided that consistent simulations exist. Computing joint likelihoods of correlated datasets, however, can be a difficult task (e.g., \cite{Nicola2020}).

Finally, we remark that our SBI-based approach can be validated, under the assumption of the adopted model, using the same catalogue generator employed to produce the training data, as demonstrated in Sec.~\ref{sec:results}. Similar catalogues can be used to quantify the goodness of fit of an analysis, as shown in Sec.~\ref{sec:goodness}. This means that a cosmological analysis can be fully carried out without the need to resort to explicit likelihoods at any point. We note that cluster number count likelihoods are often validated with synthetic catalogues as well (e.g., \cite{Zubeldia2019,Bocquet2024a,Zubeldia2024}), which means that a catalogue generator is often already part of a typical likelihood-based cluster cosmology inference pipeline.
 
Given the significant advantages outlined here, we believe that SBI has the potential to be transformative for galaxy cluster cosmology as a field. By significantly reducing both development and computational costs, it offers a highly promising pathway for obtaining cosmological constraints from galaxy cluster catalogues from current and upcoming experiments.

\section{Conclusion}\label{sec:conclusion}

In this work we have implemented an SBI-based approach for galaxy cluster number count cosmological inference. We have validated it in the context of the Simons Observatory, showing that it delivers unbiased cosmological constraints within the $\Lambda$CDM model, with any parameter biases constrained to below $0.1\,\sigma$. We have then applied it to the real \textit{Planck} MMF3 cosmology sample, finding cosmological constraints that are in excellent agreement -- better than $0.07\,\sigma$ -- with their likelihood-based counterparts from our own reanalysis. We expect this approach to be directly applicable to other real datasets, regardless of the nature of the mass observables, i.e., whether obtained through X-ray, optical or mm observations, and whether they are scalar or vector quantities (see \cite{Ishida2015,Tam2022,Kosiba2025,Reza2024}). This includes catalogues from other current and upcoming experiments such as \textit{eROSITA}, \textit{Euclid}, Rubin/LSST, SPT-3G, and CMB-S4. We also expect our approach to be directly applicable for constraining any extensions to the $\Lambda$CDM model to which cluster number counts are sensitive to. In particular, it could be used to set constraints on the dark energy equation of state and the sum of the neutrino masses.


As discussed in detail in Sec.~\ref{sec:case}, in galaxy cluster cosmology, an SBI-based approach like ours presents several major advantages over the traditional explicit likelihood-based approach: (i) it requires significantly less development effort; (ii) achieving high accuracy over the entire relevant parameter space is markedly simpler;
(iii) it can accommodate complicated models more easily; (iv) it allows for much greater computational parallelisation; and (v) it has the potential to incorporate sample variance due to cluster clustering in a much more straightforward way. In addition, a given analysis setup can be easily validated with the generation of a set of test catalogues, as demonstrated in Sec.~\ref{sec:results}, and the goodness of fit can be quantified using a similar set of catalogues, as shown in Sec.~\ref{sec:goodness}. Thus, as we have demonstrated for the first time, a cluster number count cosmological analysis can be fully carried out without resorting to an explicit likelihood at any stage. Given these advantages, we believe that SBI has the potential to be transformative for galaxy cluster cosmology as a field, significantly simplifying and accelerating the analysis of galaxy cluster catalogues from current and upcoming experiments.



\begin{table*}[t]
\caption{\label{table} True parameter values, priors, standard deviation of the likelihood-based analysis $\sigma_{\mathrm{like}}$, difference between the SBI-based and likelihood-based posterior means in units of $\sigma_{\mathrm{like}}$ for our reference test catalogue,  $(\mu_{\mathrm{SBI}} - \mu_{\mathrm{like}})  / \sigma_{\mathrm{like}}$, and constraints on the parameter biases, both in units of the SBI standard deviation $\sigma_{\mathrm{SBI}}$ and in percentage, for the analysis demonstration with our SO-like synthetic catalogues (see also Fig.~\ref{fig:constraints}). For all the parameters, $(\mu_{\mathrm{SBI}} - \mu_{\mathrm{like}})  / \sigma_{\mathrm{like}}$ is less than 0.2 and the parameter biases are constrained to being less than 0.1\,$\sigma$. Note that constraints on the first four parameters are data driven, while those on the remaining six parameters are prior driven.}
\begin{ruledtabular}
\begin{tabular}{ccccccc}
\vspace{-7pt}\\ 
Parameter & True value ($p_{\mathrm{true}}$) & Prior & $\sigma_{\mathrm{like}}$ & $(\mu_{\mathrm{SBI}} - \mu_{\mathrm{like}})  / \sigma_{\mathrm{like}}$ & $\langle (\mu_{\mathrm{SBI}} - p_{\mathrm{true}} ) \rangle /\langle \sigma_{\mathrm{SBI}} \rangle $ &  $\langle (\mu_{\mathrm{SBI}} - p_{\mathrm{true}} ) / p_{\mathrm{true}} \rangle$\,[\%] \\
\vspace{-7pt}\\
\hline \vspace{-7pt}
\\

 $\sigma_8$ &  0.811 &  $\mathcal{U}(0.7,0.9)$ &  0.0051 &  0.081 &  $ -0.107 \pm 0.086$ &  $ -0.064 \pm 0.051$  \\  $\omega_{\mathrm{c}}$ &  0.121 &  $\mathcal{U}(0.0804,0.204)$ &  0.0060 &  -0.163 &  $ 0.031 \pm 0.063$ &  $ 0.13 \pm 0.27$  \\  $A_{\mathrm{SZ}}$ &  -4.31 &  $\mathcal{U}(-4.41,-4.21)$ &  0.028 &  0.0102 &  $ 0.052 \pm 0.076$ &  $ -0.030 \pm -0.043$  \\  $\alpha_{\mathrm{SZ}}$ &  1.12 &  $\mathcal{U}(1,1.24)$ &  0.016 &  -0.0593 &  $ 0.10 \pm 0.10$ &  $ 0.13 \pm 0.15$  \\  $h$ &  0.674 &  $\mathcal{N}(0.674,0.01)$ &  0.010 &  -0.0867 &  $ 0.0211 \pm 0.0033$ &  $ 0.0309 \pm 0.0048$  \\  $\omega_{\mathrm{b}}$ &  0.0222 &  $\mathcal{N}(0.0222,0.00015)$ &  0.00015 &  -0.0373 &  $ -0.1039 \pm 0.0023$ &  $ -0.0689 \pm 0.0015$  \\  $n_{\mathrm{s}}$ &  0.96 &  $\mathcal{N}(0.96,0.0042)$ &  0.0042 &  -0.0603 &  $ -0.0395 \pm 0.0015$ &  $ -0.01767 \pm 0.00068$  \\  $\sigma_{\mathrm{SZ}}$ &  0.173 &  $\mathcal{N}(0.173,0.05)$ &  0.045 &  0.101 &  $ -0.0585 \pm 0.0063$ &  $ -1.64 \pm 0.18$  \\  $\beta_{\mathrm{CMBlens}}$ &  0.92 &  $\mathcal{N}(0.92,0.01)$ &  0.010 &  0.182 &  $ -0.0122 \pm 0.0048$ &  $ -0.0133 \pm 0.0052$  \\  $\sigma_{\mathrm{CMBlens}}$ &  0.22 &  $\mathcal{N}(0.22,0.05)$ &  0.050 &  0.127 &  $ -0.0417 \pm 0.0048$ &  $ -0.90 \pm 0.10$  \\

\end{tabular}
\end{ruledtabular}
\end{table*}

\begin{table*}[t]
\caption{\label{table2} Priors, SBI-based and likelihood-based posteriors (mean and standard deviation), and differences between the posterior means for the \textit{Planck} MMF3 cosmology sample for the CCCP+$H_0$+BBN prior combination. The agreement between the two sets of parameter constraints is excellent, with differences of less than 0.07\,$\sigma$ for all parameters. We emphasise that the adopted priors are chosen specifically to reproduce the official 2015 \textit{Planck} analysis of \cite{Ade2016}.
}
\begin{ruledtabular}
\begin{tabular}{ccccc}
\vspace{-7pt}\\ 
Parameter & Prior & SBI posterior mean $\mu_{\mathrm{SBI}}$ & Likelihood posterior mean $\mu_{\mathrm{like}}$ & $(\mu_{\mathrm{SBI}} - \mu_{\mathrm{like}})  / \sigma_{\mathrm{like}}$ \\
\vspace{-7pt}\\
\hline \vspace{-7pt}
\\

 $\sigma_8$ &  $\mathcal{U}(0.6,0.9)$ &  $0.756\pm 0.036$ &  $ 0.756 \pm 0.034$ &  $ 0.010$  \\  $\Omega_{\mathrm{m}}$ &  $\mathcal{U}(0.2,0.5)$ &  $0.342\pm 0.035$ &  $ 0.341 \pm 0.034$ &  $ 0.023$ \\ $\alpha$ &  $\mathcal{U}(1,3)$ &  $2.05\pm 0.13$ &  $ 2.04 \pm 0.13$ &  $ 0.063$  \\  $\omega_{\mathrm{b}}$ &  $\mathcal{N}(0.022,0.002)$ &  $0.0220\pm 0.0020$ &  $ 0.0220 \pm 0.0020$ &  $ -0.0078$  \\  $h$ &  $\mathcal{N}(0.738,0.024)$ &  $0.740\pm 0.025$ &  $ 0.738 \pm 0.024$ &  $ 0.054$  \\  $n_{\mathrm{s}}$ &  $\mathcal{N}(0.962,0.014)$ &  $0.962\pm 0.014$ &  $ 0.963 \pm 0.014$ &  $ -0.015$  \\  $1-b$ &  $\mathcal{N}(0.78,0.092)$ &  $0.771\pm 0.096$ &  $ 0.769 \pm 0.093$ &  $ 0.020$  \\  $\ln Y_{\star}$ &  $\mathcal{N}(-0.19,0.02)$ &  $-0.190\pm 0.020$ &  $ -0.190 \pm 0.020$ &  $ 0.048$  \\  $\sigma_{\ln Y}$ &  $\mathcal{N}(0.173,0.023)$ &  $0.177\pm 0.022$ &  $ 0.172 \pm 0.023$ &  $ 0.20$  \\

\end{tabular}
\end{ruledtabular}
\end{table*}

\begin{acknowledgments}
The authors would like to thank Boryana Hadzhiyska, Licong Xu, Peter Melchior, Joseph J.\ Mohr, Sebastian Bocquet,  Matthias Klein, James Alvey, Aditya Singh, and Sophie Vogt for useful discussions.

\'{I}Z and AC acknowledge support from the STFC (grant numbers ST/W000977/1 and ST/X006387/1).

This work was performed using resources provided by the Cambridge Service for Data Driven Discovery (CSD3) operated by the University of Cambridge Research Computing Service (\href{csd3.cam.ac.uk}{\texttt{csd3.cam.ac.uk}}), provided by Dell EMC and Intel using Tier-2 funding from the Engineering and Physical Sciences Research Council (capital grant EP/T022159/1), and DiRAC funding from the Science and Technology Facilities Council (\href{dirac.ac.uk}{\texttt{dirac.ac.uk}}), within the DiRAC Cosmos dp002 project.

Figures~\ref{fig:constraints} and~\ref{fig:constraints_planck} were produced using the \href{https://getdist.readthedocs.io/en/latest/index.html}{\texttt{getdist}} package \cite{Lewis2019}.

\end{acknowledgments}

\appendix

\section{Parameter priors, constraints and biases}\label{appendix:a}

Table\,\ref{table} lists the true parameter values (second column), adopted priors (third column), standard deviations of the likelihood-based analysis $\sigma_{\mathrm{like}}$ (fourth column), differences between the SBI-based and likelihood-based constraints in units of $\sigma_{\mathrm{like}}$ for our reference SO-like test catalogue,  $(\mu_{\mathrm{SBI}} - \mu_{\mathrm{like}})  / \sigma_{\mathrm{like}}$ (fifth column), and constraints on the parameter biases, both in units of the SBI standard deviation $\sigma_{\mathrm{SBI}}$ and in percentage (sixth and seventh columns, respectively), for the analysis demonstration with our SO-like synthetic catalogues. Notably, $(\mu_{\mathrm{SBI}} - \mu_{\mathrm{like}})$ is less than $0.2\sigma_{\mathrm{like}}$ for all parameters, the largest difference being at $-0.163\,\sigma_{\mathrm{like}}$ for $\omega_{\mathrm{c}}$. We regard these differences as negligible. Furthermore, all the biases are less than $0.1\,\sigma$. 

We note that all the parameters on which a Gaussian prior is imposed are prior driven, their likelihood-based standard deviations being equal to those of the corresponding priors. As a consequence, the statistical errors on their biases are very small, as the peak of the posterior is located very close to the peak of the prior in all realizations. The remaining four parameters, on the other hand, are data driven, with posterior standard deviations significantly smaller than those of the prior and with relatively looser constraints on their biases. These bias constraints, however, are still at the $0.1\,\sigma$ level or better.

In addition, Table\,\ref{table2} shows the priors, the SBI-based and likelihood-based posteriors (mean and standard deviation), and the differences between the posteriors for the real \textit{Planck} MMF3 cosmology sample for the CCCP+$H_0$+BBN prior combination. There is excellent agreement between the two sets of parameter constraints, with differences of less than 0.07\,$\sigma$ for all parameters. We stress that the adopted priors are chosen specifically to reproduce the official 2015 \textit{Planck} analysis of \cite{Ade2016}.

We note that $\sigma_8$, $\Omega_{\mathrm{m}}$, and $\alpha$ are data driven, with the remaining six parameters being prior driven. 

\section{Time performance: SBI vs likelihood}\label{appendix:b}

We first report the total CPU and wall-clock times required to perform the analyses for our SO-like setup. At about 12\,s per likelihood evaluation, it took 7.5\,days on one node with 76\,CPUs for 4 MCMC chains to produce a total of 82\,030 accepted samples, at which we stopped. This constitutes a total wall-clock time of 180\,h and a total CPU time of 13\,680\,h for the likelihood-based analysis. On the other hand, we used eight nodes, each with 76\,CPUs, to generate our 10\,000 training catalogues for the SBI analysis. Due to memory constraints, only one in eight CPUs were used in practice. At an average of 441\,s per catalogue (obtained by averaging over 10 randomly-chosen catalogues), this leads to a total wall-clock time of 16.1\,h and a total CPU time of 9\,807\,h. We note that the time taken to train the posterior estimate adds a negligible amount of time to this total (around $1$--30\,min on one CPU, depending on the number of simulations and learning rate).

That is, the total CPU times are comparable for both methods, with the SBI-based approach being 1.4 times cheaper than its likelihood-based counterpart. However, in terms of wall-clock time, the SBI-based approach is faster by a factor of 11.2, which constitutes an impressive mark. With enough CPUs, the wall-clock time of the SBI-based approach could be brought down to 441\,s (in addition to the few minutes required to train the posterior estimate). This is in stark contrast with the likelihood-based approach, for which such extensive parallalelisation is not possible due to the nature of MCMC methods.

Finally, we note that the time to generate the synthetic catalogues is dominated by the process of drawing mass--redshift pairs from the halo mass function. As a consequence, it scales very mildly with the nature of the cluster-observable model and the number of mass observables, something that is rarely the case for a cluster number-count likelihood. We have considered two extensions to our SO-like setup: (i) the addition of correlated intrinsic scatter between the two mass observables $q_{\mathrm{obs}}$ and $p_{\mathrm{obs}}$; and (ii) the addition of a third mass observable, which we just take to be a copy of $q_{\mathrm{obs}}$. Neither extensions lead to a statistically significant increase in the time to generate the catalogues.

\bibliographystyle{apsrev4-2}
\bibliography{references}

\end{document}